\documentclass{PoS}

\newcommand{\Chooz}{LNCA Underground Laboratory, CNRS/IN2P3 - CEA, Chooz, France}
\newcommand{\LAL}{Linear Accelerator Laboratory, CNRS/IN2P3, Universit\'e Paris-Saclay, Orsay, France}
\newcommand{\n}{{$\nu$}}
\newcommand{\Dm}{$ \Delta m^2 $}
\newcommand{\dm}{$ \delta m^2 $}
\newcommand{\angA}{$\theta _{13} $}
\newcommand{\angB}{$\theta _{12} $}
\newcommand{\angC}{$\theta _{23} $}

\title{Possible Precise Neutrino Unitarity?}

\ShortTitle{Possible Precise Neutrino Unitarity?}

\author{\speaker{Anatael Cabrera}\thanks{The author would like thank the LiquidO collaboration for ongoing fruitful discussion.}\\
	CNRS-IN2P3\\
        	\LAL\\
        	\Chooz\\
        	E-mail: \email{anatael@in2p3.fr}}

\abstract{
	The exploration of the \emph{Standard Model} (SM) leptonic mixing has been led by the study of the \emph{neutrino} (\n) \emph{oscillations} phenomenon, whose discovery was acknowledged by the 2015 Nobel prize in physics.
	Half a century of experimental and theoretical effort has established and demonstrated consistency with the 3\n\ model and with the so far SM three family evidence.
	While no direct significant manifestation for physics beyond the Standard Model (BSM) has been found, the SM is known not to suffice to explain all today's observed phenomenology.
	In the forthcoming decade, most oscillation parameters are expected to yield sub-percent precision.
	Such a knowledge opens the possibility to experimentally test for BSM manifestation(s) via the direct and competitive exploration of the {\bf PMNS matrix unitarity} for the first time.
	Any significant deviation might, in turn, evidence the existence of non-standard states (i.e. new neutrino) and/or interactions, thus allowing for direct discovery potential.
	Even if no deviations were found, the PMNS matrix structure, very different from its CKM counterpart, is of fundamental importance to our understanding of the leptonic flavour sector.
	In this document, we shall briefly review today's PMNS unitarity status in the context of existing and future particle physics programme within the next decade.
	We identify the possible need for a missing experiment(s).
	One such a case maybe a hypothetical {\bf Super Chooz}, employing the novel LiquidO technology, to address both directly sensitivity to the unitarity and unique impact to the exploration of the neutrino oscillation phenomena.
	Such a program is expected to additionally and coherently reinforce the physics of all currently planned experiments via indirect information aiding both the CP violation and mass ordering forthcoming measurements.
}

\FullConference{
	European Physical Society Conference on High Energy Physics - EPS-HEP2019 -\\
	10-17 July, 2019\\
	Ghent, Belgium}

\begin{document}

%%%%%%%%%%%%%%%%%%%%%%%%%%%%%%%%%
%\section{The Neutrino Oscillation Status}
%%%%%%%%%%%%%%%%%%%%%%%%%%%%%%%%%

\noindent
Today's neutrino oscillation experimental evidence is consistent with a 3\n\ framework~\cite{Ref_3nuGlobalAnalysis}.
This is in agreement with the observed three families of charged fermions making part of the \emph{Standard Model of Particle Physics} ({\bf SM}).
While few inconclusive indications for possible discrepancy to the 3\n\ framework have been reported, intense exploration has cornered the remaining solution phase-space to marginal region(s)~\cite{Ref_ReviewSterile} -- still not fully ruled out.
%Such discrepancies remain of limited in significance and/or linked to possible experimental issues not fully ruled out.
%Understanding them remains an important issue for future experiments.
Thus, unambiguous manifestation of physics beyond the Standard Model ({\bf BSM}) remains elusive.
%is important.
%unresolved 

%Thus, such discrepancies are generally associated to possible experimental issues still to be resolved.
Since \n\ oscillation is the macroscopic manifestation of the quantum interference of neutrino mass states during their propagation and the mixing among mass ($\nu_1$, $\nu_2$, $\nu_3$) and weak-flavour ($\nu_e$, $\nu_\mu$, $\nu_\tau$) eigenstates, the entire phenomenon is characterised in terms of 
two \emph{mass squared difference} (\dm\ and \Dm)\footnote{\dm\ and \Dm\ provide notation for the so called ``solar'' ($\Delta m_{12}^2$) and ``atmospheric'' ($\Delta m_{23}^2$ or $\Delta m_{13}^2$) cases.} 
and 
three \emph{mixing angles} (\angA, \angB, \angC), embedded in the 3$\times$3 {\bf PMNS} mixing matrix.
This is the {\bf CKM} quark counterpart.
This simplified parametrisation relies on a critical assumption: the PMNS matrix is \emph{unitary} (labelled $U$).
This same condition allows for a complex phase leading to CP-violation\footnote{The CPV implies different manifestation for matter and anti-matter, as observed in the 60's with quarks.} ({\bf CPV}) during mixing.
% and it is known to be a key ingredient for today's observed Universe formation.}.
%, unless CP-conservation solutions are measured.
There is no a priori prediction for any such parameters (6), so each must be measured to allow the phenomenological 3\n\ model characterisation of today's observations as well as possible searches for significant deviations between data and model, where discoveries may lay.
It is worth noticing that the 
%\underline
{experimental test of the unitarity is as important as the measurement of their derived parameters}.
%an assumption that must be explicitly tested.
% when considering hypothetical BSM physics.
However, testing the unitarity implies addressing a larger system of equations, where the \angA, \angB, \angC\ parametrisation no longer stands.
%, by definition.
%, as discussed later on in further detail.

%%%%%%%%%%%%%%%%%%%%%%%%%%%%%%%%%
\begin{table*}
\begin{center}
\begin{tabular}{|c|c|c|c|c|c|c|}
\hline
%& \multicolumn{3}{c|}{\bf knowledge as of 2020} & \multicolumn{3}{c|}{\bf expected knowledge beyond 2020}\\
%& precision (\%) & dominant & global (\%) & precision (\%) & dominant &technique\\
%\hline
%\angB & 3.0 & SNO & 2.3 & $\leq$1.0 & JUNO & \emph{reactor}\\
%\angC & 5.0 & NOvA & 2.0 & $\sim$1.0 & DUNE+HK & \emph{beam}\\
%\angA & 1.8 & DYB & 1.5 & 1.5 & DC+DYB+RENO & \emph{reactor}\\
%\dm & 2.5 & KL & 2.3 & $\leq$1.0 & JUNO & \emph{reactor}\\
%$|$\Dm$|$ & 3.0 & DYB+T2K & 1.3 & $\leq$1.0 & JUNO+DUNE+HK & \emph{reactor+beam}\\
%\hline
%$\pm$\Dm & unknown & SK & @\,3$\sigma$ & measure & JUNO+DUNE+HK & \emph{reactor+beam}\\
%$\delta_{\mbox{\tiny CP}}$ & unknown & T2K & @\,2$\sigma$ & measure & DUNE+HK & \emph{beam}\\\hline
& \multicolumn{2}{c|}{\bf knowledge as of 2020} & \multicolumn{3}{c|}{\bf expected knowledge beyond 2020}\\
& dominant & precision (\%) & precision (\%) & dominant &technique\\
\hline
\angB 	& SNO & 2.3 & $\leq$1.0 & JUNO & \emph{reactor}\\
\angC 	& NOvA & 2.0 & $\sim$1.0 & DUNE+HK & \emph{beam}\\
\angA 	& DYB & 1.5 & 1.5 & DC+DYB+RENO & \emph{reactor}\\
\dm 	& KL & 2.3 & $\leq$1.0 & JUNO & \emph{reactor}\\
$|$\Dm$|$ & DYB+T2K & 1.3 & $\leq$1.0 & JUNO+DUNE+HK & \emph{reactor+beam}\\
\hline
$\pm$\Dm & SK & unknown & -- & JUNO+DUNE+HK & \emph{reactor+beam}\\
$\delta_{\mbox{\tiny CP}}$ & T2K & unknown  & -- & DUNE+HK & \emph{beam}\\\hline
\end{tabular}
\end{center}
\caption{\small{\bf Neutrino Oscillation Knowledge.}
	As of 2019, current and predicted knowledge on 3\,\n\ oscillation model is summarised in terms of the precision per parameter.
	The different columns show
		today's single experiment precision,
		dominant experiment, 
		today's global precision (NuFit\,4.0),
		predicted precision 
		and 
		best experiment along with the dominant technique used.
	The entire neutrino oscillation sector will be characterised using reactors and beams.
	This is not surprising since such man-made \n's are best controlled in terms of baseline and systematics, as compared to atmospheric and solar \n's. 
	\angB\ and \angC\ will be largely improved by JUNO and DUNE+HK, respectively.
	JUNO will pioneer the sub-percent precision in the field.
	Interestingly, there is no foreseen capability to improve today's DC+DYB+RENO precision on \angA, whose knowledge will go from today's best to future worst, unless a dedicated experiment is proposed.
	%This topic will be addressed later on in the text.
	\dm\ will be dominated by JUNO while \Dm\ will be constraints by both JUNO and DUNE+HK.
	The unknown mass ordering will be addressed mainly JUNO and DUNE using vacuum oscillations and matter effects, respectively.
	Global data analysis suggests a possible favoured normal ordering solution at $\sim$3$\sigma$'s, dominated Super-Kamiokande~\cite{Ref_SK} (SK) data.
	Any deviation between JUNO and DUNE would be of great interest.
	The unknown $\delta_{\mbox{\tiny CP}}$ depends on DUNE+HK.
	Global data, dominated by T2K~\cite{Ref_T2K}, disfavours CP conservation (0 or $\pi$ solutions) at $\sim$2$\sigma$'s.
%	Atmospheric neutrino based experiments such as IceCube~\cite{Ref_IceCube} and ORCA~\cite{Ref_ORCA} are not expected to lead the ultimate precision according to their published sensitivities.
	Despite a key role in the intermediate time scale, atmospheric neutrino experiments such as IceCube~\cite{Ref_IceCube} and ORCA~\cite{Ref_ORCA} are not expected today to lead the ultimate precision by 2030.
	%Their impact is intermediate.
	%These experiments however might lead the sensitivity temporarily, if timely enough.
	\vspace{-1.0cm}
}
\label{Tab_NONow}
\end{table*}
%%%%%%%%%%%%%%%%%%%%%%%%%%%%%%%%%

For about 50 years, the experimental community has been devoted to the measurement of each of the neutrino oscillation parameters.
The key realisation was that behind the historically called \emph{solar} and \emph{atmospheric anomalies}, there is one single phenomenon: \emph{neutrino oscillation}.
However, the 2015 Nobel prize~\cite{Ref_NobelPrize} discovery acknowledgement awaited the observation of the predicted new oscillation, driven by \angA.
This was only significantly observed in 2011-2012 by Daya Bay ({\bf DYB})~\cite{Ref_DYB-I}, Double Chooz ({\bf DC})~\cite{Ref_DC-I} and {\bf RENO}~\cite{Ref_RENO-I} experiments.
Now we know neutrinos are massive even though we have not been able to measure its mass directly~\cite{Ref_PDG}.
Today's knowledge can be effectively characterised by the precision of each parameters, since no significant deviations have been found, as summarised in Table~\ref{Tab_NONow}.
While \angB\ and \angC\ are large, \angA\ is very small.
As of mid-2019, all parameters are known to the few percent (\textless2.5\%) upon combining all experiments data.
Two major unknowns remain:
%a)
\emph{atmospheric mass ordering}\footnote{This stands as the sign of \Dm\ unknown since mainly vacuum oscillation has been used to measure it. The sign of \dm\ is known due to matter dominated enhanced oscillations in the core of the sun.}
% as measured by solar experiments, where SNO  and SK experiments are dominant.}
and
%b)
the \emph{CPV phase}.
There is preliminary evidence~\cite{Ref_3nuGlobalAnalysis} amounting that
a) normal mass ordering may be favoured at $\sim$3$\sigma$'s 
and 
b) CP conservation may be disfavoured at $\sim$2$\sigma$'s.
% (maximal CP violation suggested).
%The former suggests no bounds in the phase-space to be explored by future $\beta\beta$ searches experiments.
%, as there may not be any \emph{inverted ordering bound}.
Despite major success, today's precision is still limited to address the PMNS unitarity competitively; i.e. sub-percent level.
%that might manifest.
%Although unknown, the interest is expected to raise upon $\leq$1\% precision.

In the first half of 2020 decade, the sub-percent precision regime is expected.
This will start with measurements of \angB\ and \dm\ by {\bf JUNO}~\cite{Ref_JUNO}, based in China. 
%These terms were the formerly called ``solar'' parameters.
{\bf DUNE}~\cite{Ref_DUNE} and {\bf HK}~\cite{Ref_HK}, based in USA and Japan, respectively, are expected to provide the ultimate knowledge on \angC\ during the second half of the decade.
The knowledge of \Dm, including the mass ordering resolution, is expected to be led by both JUNO and DUNE using complementary vacuum and matter effects approaches, respectively.
Surprisingly, no experiment is able to significantly improve today's \angA\ precision (1.5\%), while all experiments depend strategically on it for both CPV and mass ordering measurements.
Our \angA\ knowledge remains dominated by the aforementioned 2010 reactor data~\cite{Ref_LastDYB, Ref_DC-IV, Ref_LastRENO}.
By 2030, only experiments relying on artificially produced neutrinos (i.e. reactors and beams) will dominate the ultimate neutrino oscillation knowledge, as described in Table~\ref{Tab_NONow}.
Thus, beyond 2020, the field is expected to be shaped by a few large (or huge) experiments with the highest budgets and largest (\textgreater500 scientists) collaboration per experiment in the history of neutrino research.
%The CNRS/IN2P3 has participation to all, JUNO, DUNE and HK experiments, even if HK remains to be approved.
It is worth noticing that no major neutrino oscillation experiment is envisaged to be based in Europe in the next decade.
% or more.

In summary, upon the decade 2020-2030, the field will be reaching an overall sub-percent precision in all mixing angles, except for \angA.
The unknown mass ordering and CPV are expected to be measured by 2030 with today's data already allowing some hinted solutions at a few $\sigma$ level.
Hence, we will have all (6) parameters known by 2030.
%Some may wonder \emph{have we reached the practical end of this research line?}
%!!!!
Since, arguably, it is difficult to imagine to go larger than JUNO+DUNE+HK experiments,
%, implying a world scale effort similar to the LHC program, 
%we must therefore exhaustively ensure that {we have all needed for the most pertinent phase-space to be explored in the next decade}.
%In order words, we must ensure to be able to squeeze the most from those billions-worth of data by 2030.
%Addressing this includes, most importantly, considering the foreseeable field landscape upon the results by all the forthcoming experiments.
%!!!!!
%Otherwise said, 
so we must ensure that we are not missing anything compromising our ability to challenge the SM, thus maximising our best sensitivity to BSM possible manifestation(s), where discovery potential may be.
This reflection is be addressed timely since each step in the field is currently implying decades (preparation and data-taking) and the subsequent resources.
Indeed, this reflection is main the motivation of this document.
%, where the potential for a missing piece is preliminary identified and still under study.
%
% in the limits of feasibility.
%Indeed, addressing this reflection -- to some extent -- is the motivation of this document, where the potential for a missing piece is preliminary identified and still under study.
%might appear.

%%%%%%%%%%%%%%%%%%%%%%%%%%%%%%%%%
\section*{The PMNS Structure \& Unitarity}
%%%%%%%%%%%%%%%%%%%%%%%%%%%%%%%%%

\noindent
This is one of the most critical questions to the field -- arguably as important any consequent parameter.
Let us consider the CPV phase for instances.
% -- arguably as important as the establishment of CPV in neutrino oscillations.
% for which both DUNE and HK huge experiments are under consideration.
It is interesting to note the CPV is expected within the neutrino oscillation framework (i.e. PMNS can be complex), while there is no established model behind the violation of unitarity.
Instead, there is no SM prediction for the CPV phase value, while unitary enjoys an accurate prediction.
% of the symmetry.
% by its definition.
Hence, the unitary exploration benefits from direct discovery potential in a model-less framework exploiting an accurate prediction to identify deviations.
More, addressing unitary 
%also relies allows for the ultimate level of exploration of the PMNS related physics, thus a 
is complementary to today's measurements of each parameter, regardless of the overall PMNS structure.
% or pattern.
We shall below summarise, within the limitations of today's uncertainties, the main features of the PMNS matrix, as illustrated in Fig.\ref{Fig_PMNS}.
Its structure offers some interesting features worth some intriguing questions:

\begin{description}

\item[Why is PMNS non-diagonal?] 
	Unlike the CKM, almost diagonal, thus leading to minimal mixing in quarks, the PMNS is largely non-diagonal.
	This means its ``off-diagonal'' terms are large, as shown in Fig~\ref{Fig_PMNS}.
	This implies that whatever BSM theory may stand behind the SM effective manifestation, the predicted flavour sector may be largely different for leptons and quarks.
	%This is likely an important constraint for BSM modelling.

	It is striking to note that \angA\ is the most peculiar, as it is very small and drives the value of $U_{e3}$.
	Again, a possible hint from Nature suggesting that we ought to measure \angA\ with the highest possible precision, as it might be key to understand the leptonic flavour sector.
	Ironically, no experiment today can improve 2010's results.
	Worse, there is up to now no experimental method known to be able to challenge those results.
	A new approach is highlighted below.

%%%%%%%%%%%%%%%%%%%%%%%%%%%%%%%%%
\begin{figure}[h]
	\centering
	\includegraphics[scale=0.14]{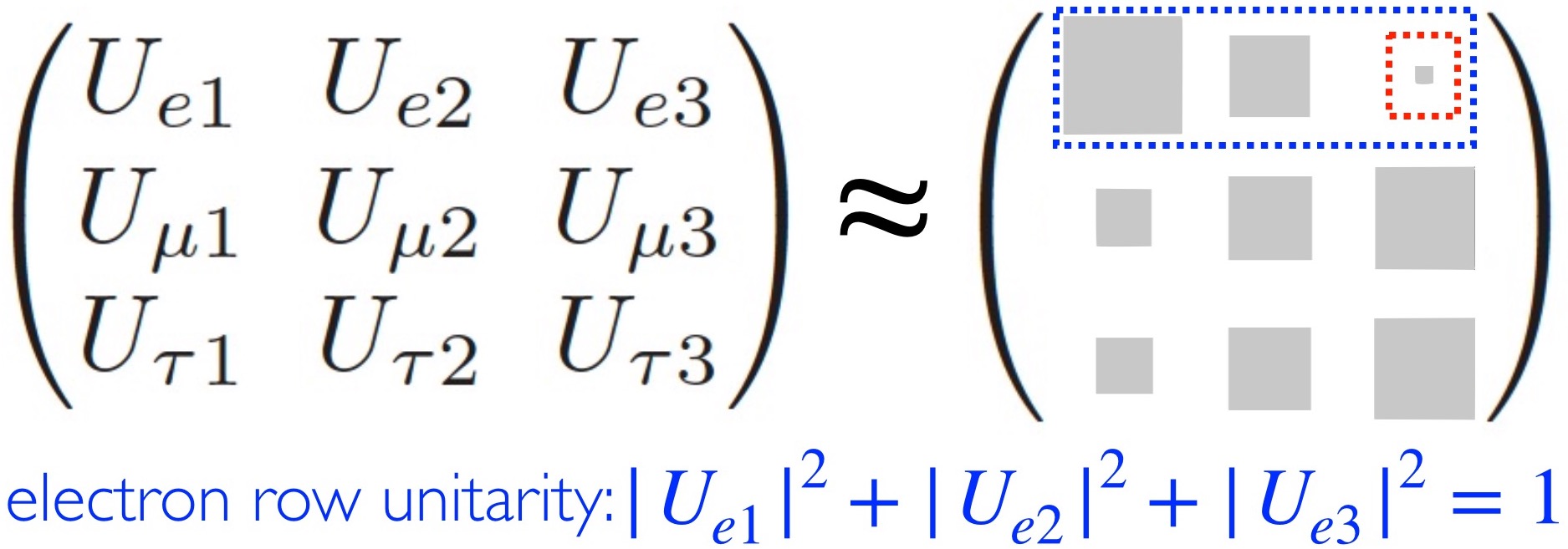}
	\caption{ \small 
	{\bf The PMNS Neutrino Mixing Matrix.}
	The non-diagonal structure and the smallness of the $U_{e3}$ term (circled in red) are among the main features of the PMNS matrix, as illustrated.
	$U_{e3}$ corresponds to \angA, if unitary.
	The overall PMNS unitarity test could be reduced to test the unitarity of the rows, where the most sensitive test today arises from the \emph{electron row} (circled in blue).
	}
	\label{Fig_PMNS}
\end{figure}
%%%%%%%%%%%%%%%%%%%%%%%%%%%%%%%%%

\item[Why is PMNS' $J$ so large?] 
	The PMNS Jarkslog invariant (factorising out the CPV phase $\sin{(\delta)}$ term) is order $\sim$10$^{-2}$, which is much larger than that of the CKM counterpart ($\sim$10$^{-5}$).
	%; order $\sim$10$^{-5}$.
	This suggests that if the CP was violated ($\sin{\delta}\neq0$), the expected CPV amplitude could be large.
	%might be the largest known.
	% source of CP violation.
	This is an appealing scenario as much CPV is needed to explain the observed matter to anti-matter asymmetry in the universe
	-- orders of magnitude more compared to the CPV embedded in the CKM.
	%~\cite{Ref_MatterAsymmetry}
	%is many orders of magnitude too large compared to that embedded in the CKM.
	%However, as of today, there is no evident relation between this CPV and the that needed to yield such a matter asymmetry upon the Big Bang.
	%The solution to this quest goes beyond the reach of the SM.
	% as known via today's physics.
	%It is however beyond today's SM theory to reach the relation between light neutrino CP violation and possible bario-genesis asymmetry mechanisms.

\item[Is PMNS unitary?] 
	As highlighted above, this is likely to be the ultimate and most challenging question that the neutrino oscillation framework might allow us to explore.
	We shall address the possible implications and today's knowledge status below.

\end{description}

\noindent
The PMNS structure largely differs from the CKM.
Hence, their nature may likely be different, although unknown.
For some, the PMNS bizarreness (compared to CKM) might indicate that its most precise exploration and scrutiny is one of the best ways to challenge the SM.
%Of course, we do not know for sure.
It would not be the first time neutrinos proved our best probe to BSM phenomenology.
One of the latest modifications in the SM was the introduction of the phenomenology of massive neutrinos, as inferred from neutrino oscillations, although the absolute scale of their lightness remains a challenging mystery.

To address the PMNS unitarity, we need an overall sub-percent mixing precision.
The results from JUNO, DUNE and HK are therefore critical.
However, while necessary, those are not sufficient conditions to yield the deepest insight.
Testing for the PMNS unitarity implies abandoning the three mixing angles (\angA, \angB, \angC) approximation.
Hence, the equations must be expressed in terms of their $U_{ij}$ terms upon imposing the unitary condition (i.e. $U U^\dagger = I$).
This translates experimentally into constraining more equations.
%However, the number of independent observables measured by the different experiments is not scaling alike.
So, {to test unitary to the percent level implies the need for the above described increase in precision but also additional measurements}.
This is described below.
Indeed, only within the 2020 decade, the field is nearing a competitive level of precision. 
%world-wide.
%; hopefully within 
%the necessary precision 
%A question remains: would  level be possible?
The reward of addressing this question is remarkable: 
{any significant evidence for unitarity violation implies the manifestation, and thus discovery, of non-standard neutrino states and\,/\,or interactions}~\cite{Ref_Hiroshi-U}. 
%Ref_NonUnitarity-Review}.
%Either way (or both), this would imply a discovery for which no BSM model whatsoever exist.
Non-standard interactions ({\bf NSI})~\cite{Ref_NSI-Review} stand for deviations during interaction and/or propagation of neutrinos.
%from the standard V-A weak interaction model for neutrinos.
This implies direct sensitivity to BSM physics manifestation despite lacking the model behind.
%, as opposed to measuring parameters predicted in an existing model. 
Given the stunning prediction power demonstrated by the SM to all so far tested observables, such those proved at the LHC, cosmology, etc; there is a diminishing phase space for direct access to discoveries in particle physics with today's technology.
Hence, testing the PMNS unitary is indeed a compelling and unique opportunity.
%In addition, by doing this we will coherently address all SM known quantities with maximal complementary.

%%%%%%%%%%%%%%%%%%%%%%%%%%%%%%%%%
\section*{The PMNS Unitarity Test Strategy}
%%%%%%%%%%%%%%%%%%%%%%%%%%%%%%%%%

\noindent
Solving the unitary condition ($U U^\dagger = I$) leads to many equations~\cite{Ref_Vogle-U}.
Some are equivalent to testing the ``closure of triangles'', as practiced in the CKM case, should the CP violation be known.
% for neutrinos.
Since, the neutrino CPV phase is still unknown, the PMNS unitary condition can be tested today via the derived $|U_{l1}|^2+|U_{l2}|^2+|U_{l3}|^2=1$ condition, with $l = e,\mu,\tau$.
These equations test the unitarity of each matrix row.
Only the $e$ and $\mu$ are considered since $\tau$ related oscillations are less constrained.
%due to the experimental challenges behind its measurement in neutrino detectors.
%, such as OPERA.
In fact, the most stringent constraint arises from the \emph{electron-row unitarity} ({\bf ERU})\footnote{The \emph{$\mu$-row} case precision is limited by experimental uncertainties such the absolute flux (typically \textless10\% for beams), the unresolved atmospheric mass ordering and the ``octant'' ambiguity due to the almost maximal value of \angC.} (or top row) leading to  
the $|U_{e1}|^2+|U_{e2}|^2+|U_{e3}|^2=1$ accurate condition.
If unitarity held, this row depends only on \angA\ and \angB.
%, which are much better known than \angC.
%This is because of the so called \emph{octant ambiguity} causes a large uncertainty due to the maximal (or almost) value of \angC.
Hence, any experiments with the ability to constrain \angA\ and/or \angB\ is of likely direct impact.
% to test the \emph{electron row .
ERU is today the only direct and most precise access to unitarity~\cite{Ref_Hiroshi-U,Ref_Vogle-U}.
% -- likely even by 2030.
% running up to 2030.
%\footnote{Other interesting and complementary tests become available with higher precision beam appearance and disappearance data.}.
% via the TR.

This is excellent news for JUNO whose highest sensitivity to \angB\ (and also \dm)
%\footnote{A team led by CNRS/IN2P3 members (LAL/SUBATECH) have developed a dedicated strategy for the measurement of \angB\ and \dm using a dual-readout approach. This approach is expected to yield novel and unique redundancy proposed originally by the CNRS/IN2P3 team.} 
unprecedentedly grants some of the necessary sub-percent precision to test ERU.
Indeed, JUNO is one of  the most important experiment in the unitarity quest~\cite{Ref_JUNO}.
However, that is not good enough.
Since, it is difficult to foresee any improvement on JUNO -- even in the far future -- we need other high precision measurements elsewhere.
Unfortunately, the sensitivity on \angA\ appears not improvable in foreseeable future.
Only DUNE, at best, might reach a similar precision as of today.
%, as highlighted before.
%Due to the low energy of reactor and solar neutrinos, only disappareanece experiments are possible.
As discussed in~\cite{Ref_Hiroshi-U,Ref_Vogle-U}, testing for ERU implies several experimental constraints, here highlighted:

\begin{description}

\item[Via \dm\ Oscillations (i.e. \angB, if unitary):]
	JUNO measures $P(\bar\nu_e \to \bar\nu_e)$ with reactor neutrinos over a $\sim$50\,km baseline.
	Also, solar neutrinos have key information by probing $P(\nu_e \to \nu_e)$ in the core of the sun via matter effects.
	Today's best constraints come from SNO and SK.
	There is no dedicated solar experiment foreseen, although JUNO has some sensitivity.
	%, beyond some marginal sensitivity by JUNO.

\item[Via \Dm\ Oscillations (i.e. \angA, if unitary):]
	again reactor experiments, like DC and DYB, had measured $P(\bar\nu_e \to \bar\nu_e)$ at the baseline of order 1\,km.
	%This remains a key ingredient.
	There is however no copious known $\nu_e$ source
	%\footnote{The only possible exception is $\pi$/$\mu$ decay-at-rest using an accelerator beam-dump or alike. However, this suffers other complications/limitations.} 
	capable of addressing $P(\nu_e \to \nu_e)$ precisely enough with a compatible L/E ratio.

\end{description}

\noindent
Although not listed explicitly above, the {\bf \emph{absolute flux} knowledge} is of critical impact to test ERU~\cite{Ref_Hiroshi-U,Ref_Vogle-U}.
%knowledge of the mass ordering is beneficial and that of the 
%This provides unique insight over possible impact of other possible non-standard oscillations affecting the unitary test.
%When  testing for unitarity, one cannot assume that the non-oscillation spectrum probability 1 without constraining to some range of phase-space.
However, the control of the absolute flux uncertainties is experimentally very challenging.
This is indeed why many neutrino oscillation experiments use multi-detectors to bypass absolute systematics, as opposed to the simpler relative systematic basis.
This way, for example, reactor experiments systematics can be controlled to the few per mille level while the absolute is controlled to order a few \% at best.
Worse, reactor neutrinos have evidenced a non-understood deficit~\cite{Ref_RAA} (2011) and spectral distorsion~\cite{Ref_DC-III} (2014) relative to ILL-data based predictions.
A hypothetical manifestation of non-standard neutrinos with \Dm at $\sim$1\,$eV^2$ had been considered.
% -- a possible non-standard neutrino state.
Today, however, such a hypothesis has lost much ground thanks to new data addressing this issue directly; i.e. weakening the hypothetical phase-space~\cite{Ref_ReviewSterile}, and/or indirect; i.e. demonstrating that the reactor prediction uncertainties are likely larger~\cite{Ref_DC-IV}.
%\footnote{The $\sim$1MeV$^2$ sterile hypothesis is disfavoured direct and indirect. Direct because XYZ. Indirectly because  XYZ.}.
Considering all those effects, today's studies~\cite{Ref_Hiroshi-U,Ref_Vogle-U} quantify that the ERU test reach a few percent (\textgreater2\%) precision, including a prospected JUNO outcome.
Hence, a dedicated experimental effort addressing the maximal sensitivity to unitarity is needed 
%involving high precision on \angA and/or the absolute flux  knowledge, 
if a sub-percent precision level is to become possible -- our goal for discussion here.
%, if possible.
%discussed next, whose ongoing quantification impact is not yet completed.
%The goal is to articulate unitarity test precision to the sub-percent level, if possible.

%%%%%%%%%%%%%%%%%%%%%%%%%%%%%%%%%
\section*{Super Chooz Project Exploration}
%%%%%%%%%%%%%%%%%%%%%%%%%%%%%%%%%

\noindent
Improving ERU test precision beyond JUNO requires
(a) a significantly better measurement of \angA\ (ideally sub-percent precision),
(b) a much better control of absolute flux 
and, possibly,
(c) a better measurement of solar neutrinos.
Unfortunately, all those items are considered today either impractical -- or even impossible -- with today's technology.

However, a new neutrino detection technology called {\bf LiquidO}~\cite{Ref_LOZ} might allow to address some of those tough challenges.
%ideally in single experimental site.
A hypothetical {\bf Super Chooz} ({\bf SC}) project has been first raised in this \emph{HEP-EPS conference}.
The project would rely on an $\sim$10\,kton LiquidO detector located in one of the existing caverns upon the final deconstruction of the former Chooz-A reactor site.
These caverns are to become available by \textgreater2025, implying minimal civil construction to reuse.
% due to existing site.
%Preliminary explorative discussion with EDF is ongoing and encouraging.
This expansion implies that the existing LNCA laboratory (Chooz) would become one of the largest underground laboratories in Europe with two of the most powerful Areva N4 reactors as source despite low overburden (300\,m water equivalent).
%~\cite{Ref_N4}
%in the world as source.
While the physics potential is still under ongoing study, the performance depends on the LiquidO detection.
%, still under intense demonstration.
The first experimental proof of principle~\cite{Ref_LOZ} using its first opaque scintillator articulation~\cite{Ref_NOWASH} has been successfully demonstrated.

%%%%%%%%%%%%%%%%%%%%%%%%%%%%%%%%%
\begin{figure}[h]
	\centering
	\includegraphics[scale=0.142]{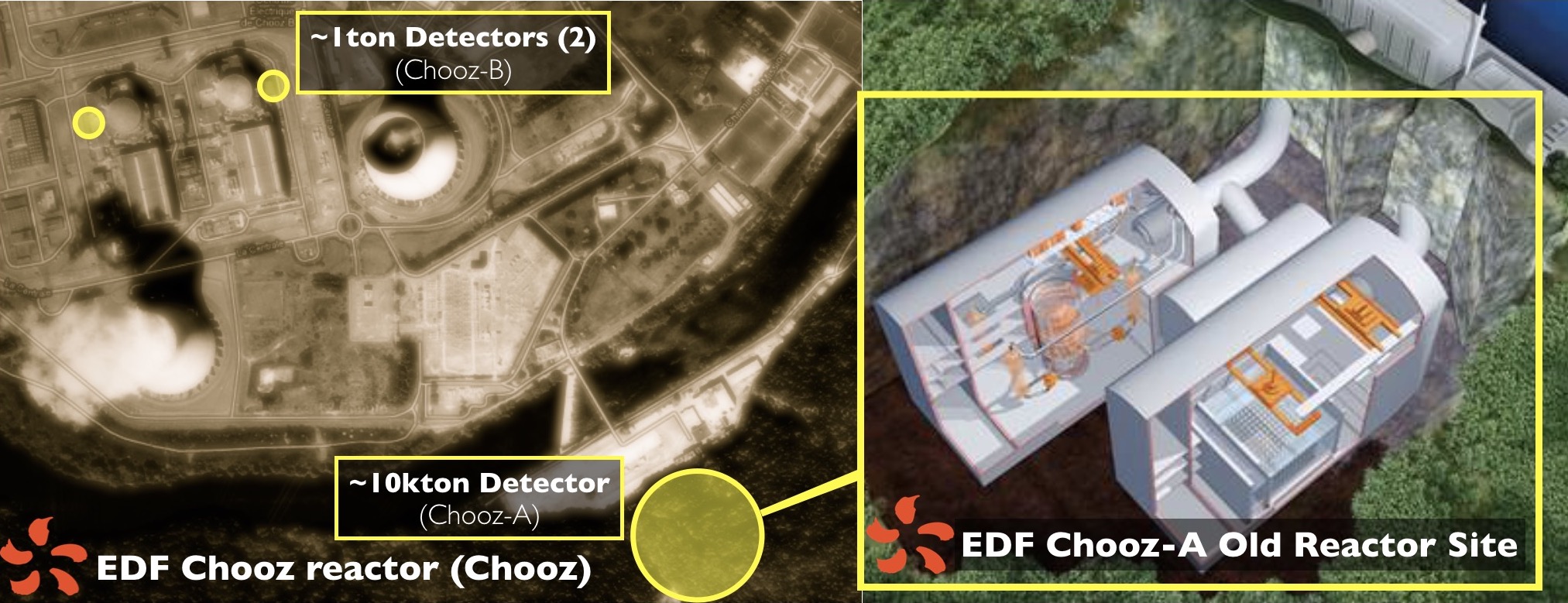}
	\caption{ \small 
	{\bf The Super Chooz Site.}
	The SC experiment relies on two very-near detectors (order 1ton each) 
	and
	one far large detector with a mass order 10\,kton.
	The multi-purpose far detector provide most of the physics programme (details in text).
	The site relies on the unique opportunity to use the former EDF Chooz-A reactor caverns for scientific purposes, thus expanding the existing LNCA laboratory up to 50\,m$^3$ volume.
	}
	\label{Fig_SC}
\end{figure}
%%%%%%%%%%%%%%%%%%%%%%%%%%%%%%%%%

Super Chooz addressees the necessary to yield a \angA\ measurement to \textless1\% precision -- publication soon.
This is a possible breakthrough since no technique so far is known to be able to reach such a precision.
This precision could further aid enhance the sensitivities of DUNE and HK on CPV, similar to today's reactor experiments have aided T2K sensitivity
and
possibly also JUNO's mass ordering. 
%are expected to improve from both such inputs.
%The quantification of this critical synergies are ongoing.
%However, again, this might not be enough improve ERU test sizeably.
%This potential drives the above mentioned detector size and the need for LiquidO technology\footnote{LiquidO has unique capability to strongly reduce cosmogenic backgrounds (event-wise ID) and control both energy and detection systematics.}.
A novel technique called \emph{reactor flux decomposition} is needed to yield total reactor flux error cancellation, as the near detector technique, a la DC or DYB, are proved insufficient.
%, which is supersede the ``iso-flux'' technique used by DC.
%In fact, the near detector technique, a la DC, has been demonstrated to be totally deprecated.
Our preliminary studies suggest the world best precision on both \angA\ and \Dm via shape extraction may be possible. 
Hence 2 (out of 6) of the parameters listed in Table~\ref{Tab_NONow} can be improved by Super Chooz, whose experimental configuration is shown in Fig~\ref{Fig_SC}.
SC might be  able to help the two other measurements needed for better unitarity precision absolute reactor flux and solar neutrinos, but this is under intense exploration still.
%Despite a likely ideal configuration, the former is likely impractical due to reactor prediction biasses.
Solar neutrino might benefit from an indium loading~\cite{Ref_LOZ} to enable unprecedented solar neutrino measurement via CC interactions, unlike the more challenging electron elastic scattering.
The main challenge here however is the control of cosmogenic backgrounds due to the lower overburden, while precise (mm scale) $\mu$ precise tracking might open for unprecedented tagging of the correlation between the primary $\mu$ and the spallation products.

%\begin{description}
%
%\item[Reactor Neutrino Absolute Flux.] 
%	The reactor flux decomposition technique alluded before implies the need for small LiquidO detectors at $\sim$20\,m from each reactor~\cite{Ref_LO@EPS}. 
%	They could provide the most precise reactor flux rate measurements -- ongoing study.
%	New techniques are likely needed and under consideration.
%	This is complementary to the JUNO~\cite{Ref_TAO} best reactor spectral reference using the dedicated TAO detector.
%	%provided by TAO detector in JUNO.
%
%\item[Solar Neutrino Measurement.]
%	Upon indium loading~\cite{Ref_LOZ}, SC might allow unprecedented solar neutrino measurement via CC interactions, unlike electron elastic scattering.
%	The main challenge is to be able to handle cosmogenic backgrounds due to the lower overburden.
%	LiquidO's $\mu$ precise tracking is expected to allow for unprecedented tagging between the primary $\mu$ and the spallation products.
%	% correlation tagging.
%	With JUNO's measurements of \dm\ and \angB, solar neutrinos would allow further constrain of the unitary test as well as probing both sun physics and NSI at the longest possible baselines, hence expanding SC scope to BSM searches. 
%	This scenario is also under active ongoing study.
%	%~\cite{Ref_LOS}.
%
%\end{description}

% can do even more physics thanks to the size and the powerful LiquidO technology:

%\noindent

%!!!!
%Furthermore, 
Super Chooz could also become one of the best supernova neutrino (burst \& remnant) and proton decay detectors, while complementary to other foreseen detectors.
On the supernova side, the ability for LiquidO to detect and identify $\bar\nu_e$ and $\nu_e$, upon CC interactions, allows unique capability for supernova neutrinos (\textless50\,MeV), including major background reduction and event-wise directionality.
Flavour independent NC interaction detection is also possible upon loading, as highlighted in~\cite{Ref_LOZ}.
The supernova potential remains under active ongoing study.
%~\cite{Ref_LOSN}.
On the proton decay side, LiquidO's event-wise imaging, again, allows the event-wise identification of K$^+$, $\pi^0$, $\pi^\pm$, $\mu^\pm$, etc. via their main decay mode(s).
All of those particles play a role in different proton decay modes.
LiquidO is expected thus to be one of the best proton decay searches technologies in terms of 
its highest free-proton density (normal in scintillators), 
high efficiency of detection 
and 
possible multi-decay mode sensitivity, boosted by its expected large background rejection.
This was preliminary highlighted in~\cite{Ref_LO@CERN}, but further studies are ongoing.
%, but further studies are ongoing.
%This is still in demonstration phase in the fellowing few years.

The feasibility and vast physics programme potential of a hypothetical Super Chooz is under study within the LiquidO collaboration, supported by several other cooperating institutions.
%, putting together experimentalists and phenomenologists word-wide for this exploration.
The Super Chooz project could have a unique and major potential impact to field, should the LiquidO technology performance demonstrates.
The programme is complementary to all JUNO, DUNE and HK.
%, under parallel experimental validation.
%Its feasibility and full physics programme is expected to be clarified shortly.
The main aspiration remains to articulate a comprehensive programme to tackle all the measurements needed to yield maximal PMNS unitarity precision. 
%, test with sub-percent precision.
The success of this ambitious goal is under exploration but relies also on the precision to become available in the next decade. 

\newpage
%%%%%%%%%%%%%%%%%%%%%%%%%%%%%%

\end{document}